\documentclass[runningheads,a4paper]{llncs}

\usepackage{mdframed}
\usepackage{hyperref}
\usepackage{times}
\usepackage{amsmath,amssymb, latexsym}
\usepackage[latin1]{inputenc}
\usepackage{amsmath}
\usepackage{amsfonts}
\usepackage{amssymb}
\usepackage{commands}

\usepackage{enumerate}
\def\url{}
\usepackage{url}
\DeclareUrlCommand\UScore{\urlstyle{rm}}
\usepackage{xspace}
\usepackage{epsfig}
\usepackage{booktabs}
\usepackage{ifthen}
\usepackage{siunitx}
\usepackage{etoolbox}

\usepackage{fancybox}
\makeatletter
\newenvironment{CenteredBox}{%
\begin{Sbox}}{
\end{Sbox}\centerline{\parbox{\wd\@Sbox}{\TheSbox}}}
\makeatother

\usepackage{pgfplots}
\usepackage{pgfkeys}
\pgfplotsset{width=0.38\linewidth} 
\usepgfplotslibrary{groupplots}

\makeatletter
\pgfplotsset{
    groupplot xlabel/.initial={},
    every groupplot x label/.style={
        at={($({group c1r\pgfplots@group@rows.west}|-{group c1r\pgfplots@group@rows.outer south})!0.5!({group c\pgfplots@group@columns r\pgfplots@group@rows.east}|-{group c\pgfplots@group@columns r\pgfplots@group@rows.outer south})$)},
        anchor=north,
    },
    groupplot ylabel/.initial={},
    every groupplot y label/.style={
            rotate=90,
        at={($({group c1r1.north}-|{group c1r1.outer
west})!0.5!({group c1r\pgfplots@group@rows.south}-|{group c1r\pgfplots@group@rows.outer west})$)},
        anchor=south
    },
    execute at end groupplot/.code={%
      \node [/pgfplots/every groupplot x label]
{\pgfkeysvalueof{/pgfplots/groupplot xlabel}};  
      \node [/pgfplots/every groupplot y label] 
{\pgfkeysvalueof{/pgfplots/groupplot ylabel}};  
    },
    group/only outer labels/.style =
{
group/every plot/.code = {%
    \ifnum\pgfplots@group@current@row=\pgfplots@group@rows\else%
        \pgfkeys{xticklabels = {}, xlabel = {}}\fi%
    \ifnum\pgfplots@group@current@column=1\else%
        \pgfkeys{yticklabels = {}, ylabel = {}}\fi%
}
}
}

\def\endpgfplots@environment@groupplot{%
    \endpgfplots@environment@opt%
    \pgfkeys{/pgfplots/execute at end groupplot}%
    \endgroup%
}
\makeatother

\newcommand{\includeProof}[1]{
  \ifthenelse{\equal{\isTechReport}{true}}{
    #1
  }{
  }
}

\def\@envspa{\hspace{0.3em}}
\def\@sa{\hspace{-0.2em}}
\def\@sb{\hspace{0.5em}}
\def\@sc{\hspace{-0.1em}}

{\itshape}{}

\newcommand{\isTechReport}{false} 
\newcommand\includeTechRep[1]{%
  \ifthenelse{\equal{\isTechReport}{true}}
    {{#1}}
    {\ignorespaces}
\xspace}

\usepackage{listings}
\usepackage{color}

\ifdefined\withcolor
	\definecolor{haskellblue}{rgb}{0.0, 0.0, 1.0}
	\definecolor{haskellstr}{rgb}{0.2, 0.2, 0.6}
	\definecolor{haskellred}{rgb}{1.0, 0.0, 0.0}
	\definecolor{gray_ulisses}{gray}{0.55}
	\definecolor{castanho_ulisses}{rgb}{0.71,0.33,0.14}
	\definecolor{preto_ulisses}{rgb}{0.41,0.20,0.04}
	\definecolor{green_ulisses}{rgb}{0.0,0.4,0.0}
\else
	\definecolor{haskellblue}{gray}{0.1}
	\definecolor{haskellstr}{gray}{0.1}
	\definecolor{haskellred}{gray}{0.1}
	\definecolor{gray_ulisses}{gray}{0.1}
	\definecolor{castanho_ulisses}{gray}{0.1}
	\definecolor{preto_ulisses}{gray}{0.1}
	\definecolor{green_ulisses}{gray}{0.1}
\fi

\def\codesize{\footnotesize}

\lstdefinelanguage{HaskellUlisses}{
	basicstyle=\codesize\ttfamily,
	sensitive=true,
	morecomment=[l][\color{gray_ulisses}\ttfamily\codesize]{--},
	morecomment=[s][\color{gray_ulisses}\ttfamily\codesize]{\{-}{-\}},
	morestring=[b]",
	stringstyle=\color{haskellstr},
	showstringspaces=false,
	numberstyle=\codesize,
	numberblanklines=true,
	showspaces=false,
	breaklines=true,
	showtabs=false,
	emph=
	{[1]
		FilePath,IOError,abs,acos,acosh,and,any,appendFile,approxRational,asTypeOf,asin,
		asinh,atan,atan2,atanh,basicIORun,break,catch,ceiling,chr,compare,concat,concatMap,
		const,cos,cosh,curry,cycle,decodeFloat,denominator,digitToInt,div,divMod,drop,
		dropWhile,either,elem,encodeFloat,enumFrom,enumFromThen,enumFromThenTo,enumFromTo,
		error,even,exp,exponent,fail,filter,flip,floatDigits,floatRadix,floatRange,floor,
		fmap,foldl,foldl1,foldr,foldr1,fromDouble,fromEnum,fromInt,fromInteger,
		fromRational,fst,gcd,getChar,getContents,getLine,head,id,inRange,index,init,intToDigit,
		interact,ioError,isAlpha,isAlphaNum,isAscii,isControl,isDenormalized,isDigit,isHexDigit,
		isIEEE,isInfinite,isLower,isNaN,isNegativeZero,isOctDigit,isPrint,isSpace,isUpper,iterate,
		last,lcm,length,lex,lexDigits,lexLitChar,lines,log,logBase,lookup,map,mapM,mapM_,max,
		maxBound,maximum,maybe,min,minBound,minimum,mod,negate,not,notElem,numerator,odd,
		or,pi,primExitWith,print,product,properFraction,putChar,putStr,putStrLn,quot,
		quotRem,range,rangeSize,read,readDec,readFile,readFloat,readHex,readIO,readInt,readList,readLitChar,
		readLn,readOct,readParen,readSigned,reads,readsPrec,realToFrac,recip,rem,repeat,replicate,
		reverse,round,scaleFloat,scanl,scanl1,scanr,scanr1,seq,sequence,sequence_,show,showChar,showInt,
		showList,showLitChar,showParen,showSigned,showString,shows,showsPrec,significand,signum,sin,
		sinh,snd,span,splitAt,sqrt,subtract,succ,sum,tail,take,takeWhile,tan,tanh,threadToIOResult,toEnum,
		toInt,toInteger,toLower,toRational,toUpper,truncate,uncurry,undefined,unlines,until,unwords,unzip,
		unzip3,userError,words,writeFile,zip,zip3,zipWith,zipWith3,listArray,doParse,for,initTo,
                create,get,set,div,rescale,add,delete,insert,prop_focus_left_master,average,best,insert,union,split,size,fromList,copy,group,good,bad,foo,explode,singleton,difference,fromJust,sort,unfold,
                target, query, decode, encode, check, refuteSMT, binder,
                subst,unapply,apply,proxy,refinement,fresh,guard,constrain,oneOf, 
                queryList,queryCtor,queryField,ctors,decodeCtor,whichOf,ctorArity,eval,
                mkCtor,gCtors,gEncode,gEncodeFields,gDecode,gDecodeFields,reproxyRep,empty,splitCtor,checkField,scanM, 
                padAverage,focusUp,execute,checkSMT,inputTypes,outputType,toReft,app
	},
	emphstyle={[1]\color{haskellblue}},
	emph=
	{[2]
		OkMap,OkRBT,OkStackSet,TTrue,Map,Bool,Char,Double,Either,Float,IO,Integer,Int,Maybe,Ordering,Rational,Ratio,ReadS,ShowS,String,Word8,Nat,Pos,Rng,Score,
                Ptr,ForeignPtr,CSize,InPacket,Tree,Prop,TreeEq,TreeLt,Vec,
                NullTerm,IncrList,DecrList,UniqList,BST,MinHeap,MaxHeap,
                PtrN,ByteStringN,ByteStringEq,VO,ByteStringsEq,ByteStringNE,OrdList,Var,RType,Constrain,Gen,Var,Proxy,SMT,Targetable,RefType,Refinement,Ctor,C1,Rep,Rec0,U1,
                GCtors,GDecode,GDecodeFields,GEncode,GEncodeFields,OrdMap,MinusKey,
                len,isBH,isBal,bh,isRB,keys,List,Sorted,RBT,Col,isBlack,OrdRBT,Set,sz,
                StackSet,NoDuplicates,Data,RBTree,XMonad,Generic,true
	},
	emphstyle={[2]\color{castanho_ulisses}},
	emph=
	{[3]
		case,class,data,deriving,do,else,if,return,def,import,in,infixl,infixr,instance,let,tmapM,for2M,forM,zipWithM,otherwise,
		module,measure,pred,predicate,of,primitive,then,type,where,lazy,throw,when
	},
	emphstyle={[3]\color{preto_ulisses}\textbf},
	emph=
	{[4]
		quot,rem,div,mod,elem,notElem,seq
	},
	emphstyle={[4]\color{castanho_ulisses}\textbf},
	emph=
	{[5]
		PS,Tip,Node,Black,Red,EQ,False,GT,Just,LT,Left,Nothing,Right,True,Show,Eq,Ord,Num,C,N,Leaf,Bin,CounterExample
	},
	emphstyle={[5]\color{green_ulisses}},
	emph=
	{[6]
		patError, irrefutPatError, nonExhaustiveGuardsError, recSelError, errorOut, 
		noMethodBinding
	},
	emphstyle={[6]\color{haskellred}}
}

\lstdefinelanguage{HaskellUlissesMath}[]{HaskellUlisses}{mathescape=true}

\lstnewenvironment{code}
{\lstset{language=HaskellUlisses}}
{}

\lstnewenvironment{mcode}
{\lstset{language=HaskellUlissesMath}}
{}

\lstnewenvironment{fcode}
{\lstset{language=HaskellUlisses, frame=L}}
{}

\lstMakeShortInline[language=HaskellUlisses]@
\lstMakeShortInline[language=HaskellUlissesMath]|

\title{Type Targeted Testing}
\author{Eric L. Seidel \and Niki Vazou \and Ranjit Jhala}
\institute{UC San Diego}

\begin{document}
\maketitle

\begin{abstract}
We present a new technique called \emph{type targeted testing}, 
which translates precise \emph{refinement types} into comprehensive 
test-suites.
The key insight behind our approach is that through the lens of SMT
solvers, refinement types can also be viewed as a high-level, 
declarative, test generation technique, wherein types are converted
to SMT queries whose models can be decoded into concrete program inputs.
Our approach enables the systematic and exhaustive testing of 
implementations from high-level declarative specifications, and 
furthermore, provides a gradual path from testing to full verification.
We have implemented our approach as a Haskell testing tool called 
\toolname, and present an evaluation that shows how \toolname can 
be used to test a wide variety of properties and how it compares 
against state-of-the-art testing approaches.
\end{abstract}


\section{Introduction}\label{sec:intro}

Should the programmer spend her time writing \emph{better types}
or \emph{thorough tests}?  
Types have long been the most pervasive means of describing the 
intended behavior of code. However, a type signature is often a 
very coarse description; the actual inputs and outputs
may be a subset of the values described by the types. 
For example, the set of ordered integer lists is a very 
sparse subset of the set of all integer lists. 
Thus, to validate functions that produce or consume such values, 
the programmer must painstakingly enumerate these values by hand 
or via ad-hoc generators for unit tests.

We present a new technique called \emph{type targeted testing}, 
abbreviated to \toolname, that enables the generation of unit
tests from precise \emph{refinement types}.
Over the last decade, various groups have shown how refinement 
types -- which compose the usual types with logical refinement predicates
that characterize the subset of actual type inhabitants -- 
can be used to specify and formally verify a wide variety 
of correctness properties of programs~\cite{pfenningxi98,Dunfield07,fstar,VazouICFP14}.
Our insight is that through the lens of SMT
solvers, refinement types can be viewed as a high-level, 
declarative, test generation technique.

\toolname tests an implementation function against a refinement 
type specification using a \emph{query-decode-check} loop.
%
First, \toolname translates the argument types into a logical
\emph{query} for which we obtain a satisfying assignment 
(or model) from the SMT solver.
Next, \toolname \emph{decodes} the SMT solver's model to obtain
concrete input values for the function.
Finally, \toolname executes the function on the inputs 
to get the corresponding output, which we \emph{check} 
belongs to the specified result type. 
If the check fails, the inputs are returned as a counterexample, 
otherwise
\toolname refutes the given model to force the SMT solver to 
return a different set of inputs. 
This process is repeated for a given number of 
iterations, or until \emph{all} inputs up to a certain size 
have been tested.

\toolname offers several benefits over other testing techniques.
Refinement types provide a succinct description of the 
input and output requirements, eliminating the need to 
enumerate individual test cases by hand or to write 
custom generators.
Furthermore, \toolname generates \emph{all} 
values (up to a given size) that inhabit a type, and thus
does not skip any corner cases that a hand-written generator 
might miss.
Finally, while the above advantages can be recovered by a brute-force
generate-and-filter approach that discards inputs that do not meet
some predicate, we show that our SMT-based method can be significantly
more efficient for enumerating valid inputs in a highly-constrained space.

\toolname paves a \emph{gradual path} from testing to verification, 
that affords several advantages over verification.
First, the programmer has an \emph{incentive} to write formal 
specifications using refinement types. \toolname provides the 
immediate gratification of an automatically generated, 
exhaustive suite of unit tests that can expose errors.
Thus, the programmer is rewarded without paying, up front, 
the extra price of annotations, hints, strengthened 
inductive invariants, or tactics needed for formally 
verifying the specification.
Second, our approach makes it possible to use refinement 
types to formally verify \emph{some} parts of the program, 
while using tests to validate other parts that may
be too difficult to verify
\toolname integrates the two modes by using refinement
types as the uniform specification mechanism. 
Functions in the verified half can be formally checked 
\emph{assuming} the functions in the tested half adhere 
to their specifications. 
We could even use refinements to generate dynamic 
contracts~\cite{Findler01} around the tested half 
if so desired.
Third, even when formally verifying the type specifications, 
the generated tests can act as valuable \emph{counterexamples} 
to help \emph{debug} the specification or implementation in 
the event that the program is rejected by the verifier.

Finally, \toolname offers several concrete advantages over previous
property-based testing techniques, which also have the potential for 
gradual verification.
First, instead of specifying properties with arbitrary code 
\cite{claessen_quickcheck:_2000,runciman_smallcheck_2008} 
which complicates the task of subsequent formal verification, 
with \toolname the properties are specified via refinement 
types, for which there are already several existing formal 
verification algorithms~\cite{VazouICFP14}.
Second, while symbolic execution tools~\cite{DART,CUTE,Veanes08} 
can generate tests from arbitrary code contracts (\eg assertions) 
we find that highly constrained inputs trigger path explosion 
which precludes the use of such tools for gradual verification.

In the rest of this paper, we start with an overview of 
how \toolname can be used and how its query-decode-check 
loop is implemented (\S~\ref{sec:overview}).
Next, we formalize a general framework for type-targeted 
testing (\S~\ref{sec:framework}) and show how it can be 
instantiated to generating tests for lists (\S~\ref{sec:list}), 
and then automatically generalized to other 
types (\S~\ref{sec:generic}).
All the benefits of \toolname come at a price; 
we are limited to properties that can be specified with 
refinement types. 
We present an empirical evaluation that shows
\toolname is efficient and expressive enough to capture 
a variety of sophisticated properties,
demonstrating that type-targeted 
testing is
a sweet spot between automatic testing 
and verification (\S~\ref{sec:evaluation}).

\section{Overview}\label{sec:overview}

We start with a series of examples pertaining to a small grading
library called @Scores@. The examples provide a bird's eye view of 
how a user interacts with \toolname, how \toolname is implemented,
and the advantages of type-based testing.

\mypara{Refinement Types}
A refinement type is one where the basic types are decorated 
with logical predicates drawn from an efficiently decidable 
theory. For example,
\begin{code}
  type Nat   = {v:Int | 0 <= v}
  type Pos   = {v:Int | 0 <  v}
  type Rng N = {v:Int | 0 <= v && v <  N}
\end{code}
are refinement types describing the set of integers that are 
non-negative, strictly positive, and in the interval @[0, N)@ 
respectively. We will also build up function and collection 
types over base refinement types like the above. 
In this paper, we will not address the issue of \emph{checking}
refinement type signatures~\cite{VazouICFP14}.
We assume the code is typechecked, \eg by GHC, against the 
standard type signatures obtained by erasing the refinements.
Instead, we focus on using the refinements to 
synthesize tests to \emph{execute} the function, and to find 
\emph{counterexamples} that violate 
the given specification.

\subsection{Testing with Types}

\mypara{Base Types}
Let us write a function @rescale@ that takes a source range @[0,r1)@, 
a target range @[0,r2)@, and a score @n@ from the source range,
and returns the linearly scaled score in the target range.
For example, @rescale 5 100 2@ should return @40@. 
Here is a first attempt at @rescale@ 
\begin{code}
  rescale :: r1:Nat -> r2:Nat -> s:Rng r1 -> Rng r2 
  rescale r1 r2 s = s * (r2 `div` r1)   
\end{code}
When we run \toolname, it immediately reports 
\begin{code}
  Found counter-example: (1, 0, 0) 
\end{code}
Indeed, @rescale 1 0 0@ results in @0@ which is not in the target 
@Rng 0@, as the latter is empty! We could fix this in various ways, 
\eg by requiring the ranges are non-empty:
\begin{code}
  rescale :: r1:Pos -> r2:Pos -> s:Rng r1 -> Rng r2 
\end{code}
Now, \toolname accepts the function and reports
\begin{code}
  OK. Passed all tests.
\end{code}
Thus, using the refinement type \emph{specification} for @rescale@, 
\toolname systematically tests the \emph{implementation} by generating 
all valid inputs (up to a given size bound) that respect the 
pre-conditions, running the function, and checking that the 
output satisfies the post-condition.
Testing against random, unconstrained inputs would be of limited value 
as the function is not designed to work on all @Int@ values. While in 
this case we could filter invalid inputs, we shall show
that \toolname can be more effective.

\mypara{Containers}
Let us suppose we have normalized all scores to be out of @100@
\begin{code}
  type Score = Rng 100
\end{code}
Next, let us write a function to compute a \emph{weighted} average 
of a list of scores.
\begin{code}
  average     :: [(Int, Score)] -> Score
  average []  = 0
  average wxs = total `div` n
    where
      total   = sum [w * x | (w, x) <- wxs ]
      n       = sum [w     | (w, _) <- wxs ]
\end{code}
It can be tricky to \emph{verify} this function as it requires non-linear reasoning
about an unbounded collection. However, we can gain a great degree of confidence by
systematically testing it using the type specification; indeed, \toolname responds:
\begin{code}
  Found counter-example: [(0,0)]
\end{code}
Clearly, an unfortunate choice of weights can trigger a divide-by-zero; we can fix 
this by requiring the weights be non-zero:
\begin{code}
  average :: [({v:Int | v /= 0}, Score)] -> Score
\end{code}
but now \toolname responds with
\begin{code}
  Found counter-example: [(-3,3),(3,0)]
\end{code}
which also triggers the divide-by-zero! We will play it safe and require positive weights,
\begin{code}
  average :: [(Pos, Score)] -> Score
\end{code}
at which point \toolname reports that all tests pass.

\mypara{Ordered Containers}
The very nature of our business requires that at the end of the day,
we order students by their scores. We can represent ordered lists by 
requiring the elements of the tail @t@ to be greater than the head @h@:
\begin{code}
data OrdList a = [] | (:) {h :: a, t :: OrdList {v:a | h <= v}}
\end{code}
Note that erasing the refinement predicates gives us plain old Haskell lists.
We can now write a function to insert a score into an ordered list:
\begin{code}
  insert :: (Ord a) => a -> OrdList a -> OrdList a 
\end{code}
\toolname automatically generates all ordered lists (up to a given size)
and executes @insert@ to check for any errors. Unlike randomized testers, 
\toolname is not thwarted by the ordering constraint, and does not require a
custom generator from the user.

\mypara{Structured Containers} 
Everyone has a few bad days. Let us write a function that takes the 
@best k@ scores for a particular student. That is, the output
must satisfy a \emph{structural} constraint -- that its size 
equals @k@. We can encode the size of a list with a logical 
measure function~\cite{VazouICFP14}:
\begin{code}
  measure len :: [a] -> Nat
  len []      = 0
  len (x:xs)  = 1 + len xs
\end{code}
Now, we can stipulate that the output indeed has @k@ scores:
\begin{code}
  best      :: k:Nat -> [Score] -> {v:[Score] | k = len v}
  best k xs = take k $ reverse $ sort xs
\end{code}
Now, \toolname quickly finds a counterexample:
\begin{code}
  Found counter-example: (2,[])
\end{code}
Of course -- we need to have at least @k@ scores to start with! 
\begin{code}
best :: k:Nat -> {v:[Score]|k <= len v} -> {v:[Score]|k = len v}
\end{code}
and now, \toolname is assuaged and reports no counterexamples.
While randomized testing would suffice for @best@, we will see 
more sophisticated structural properties such as height balancedness, 
which stymie random testers, but are easily handled by \toolname.

\mypara{Higher-order Functions} 
Perhaps instead of taking the $k$ best grades, we would like
to pad each individual grade, and, furthermore, we want to
be able to experiment with different padding functions. Let
us rewrite @average@ to take a functional argument, and
stipulate that it can only increase a @Score@.
\begin{code}
  padAverage       :: (s:Score -> {v:Score | s <= v}) 
                   -> [(Pos, Score)] -> Score
  padAverage f []  = f 0
  padAverage f wxs = total `div` n
    where
      total   = sum [w * f x | (w, x) <- wxs ]
      n       = sum [w       | (w, _) <- wxs ]
\end{code}
\toolname automatically checks that @padAverage@ is 
a safe generalization of @average@. Randomized 
testing tools can also generate functions, but those 
functions are unlikely to satisfy non-trivial constraints, 
thereby burdening the user with custom generators.

\subsection{Synthesizing Tests} 
\label{sec:synthesizing-tests}
Next, let us look under the hood to get an idea of how \toolname 
synthesizes tests from types. 
At a high-level, our strategy is to:
(1)~\emph{query}   an SMT solver for satisfying assigments to a set of logical 
                   constraints derived from the refinement type,
(2)~\emph{decode}  the model into Haskell values that are suitable inputs,
(3)~\emph{execute} the function on the decoded values to obtain the output, 
(4)~\emph{check}   that the output satisfies the output type,
(5)~\emph{refute}  the model to generate a different test, and 
repeat the above steps until all tests up to a certain size are executed.
We focus here on steps 1, 2, and 4 -- query, decode, and check -- the others are 
standard and require little explanation.

\mypara{Base Types}
Recall the initial (buggy) specification
\begin{code}
  rescale :: r1:Nat -> r2:Nat -> s:Rng r1 -> Rng r2 
\end{code}
\toolname \emph{encodes} input requirements for base types directly 
from their corresponding refinements. The constraints for multiple, 
related inputs are just the \emph{conjunction} of the constraints 
for each input. Hence, the constraint for @rescale@ is:
$$
\cstr{C_0} \defeq 0 \leq \cvar{r1} \wedge 0 \leq \cvar{r2} \wedge 0 \leq s < \cvar{r1} 
$$
In practice, $\cstr{C_0}$ will also contain conjuncts of the form $-N \leq x \leq N$ that
restrict @Int@-valued variables $x$ to be within the size bound $N$ supplied by
the user, but we will omit these throughout the paper for clarity.

Note how easy it is to capture dependencies between inputs, 
\eg that the score @s@ be in the range defined by @r1@.
On querying the SMT solver with the above, we get a model
$[\cvar{r1} \mapsto 1, \cvar{r2} \mapsto 1, \cvar{s}  \mapsto 0]$.
\toolname decodes this model and executes \hbox{@rescale 1 1 0@} to obtain the value @v = 0@.
Then, \toolname validates @v@ against the post-condition by checking 
the validity of the output type's constraint: 
$$\cvar{r2} = 1 \wedge \cvar{v} = 0 \wedge 0 \leq \cvar{v} \wedge \cvar{v} < \cvar{r2}$$
As the above is valid, \toolname moves on to generate another 
test by conjoining $\cstr{C_0}$ with a constraint that refutes 
the previous model:
$$
\cstr{C_1} \defeq \cstr{C_0} \wedge (\cvar{r1} \not = 1 \vee \cvar{r2} \not = 1 \vee \cvar{s} \not = 0)
$$
This time, the SMT solver returns a model: 
$[\cvar{r1} \mapsto 1, \cvar{r2} \mapsto 0, \cvar{s} \mapsto 0]$
which, when decoded and executed, yields the result $0$ that does \emph{not} 
inhabit the output type, and so is reported as a counterexample. 
When we fix the specification to only allow @Pos@ ranges, each test produces
a valid output, so \toolname reports that all tests pass.

\mypara{Containers}
Next, we use \toolname to test the implementation of @average@.
To do so, \toolname needs to generate Haskell lists with the appropriate constraints.
Since each list is recursively 
either ``nil'' 
or ``cons'', 
\toolname generates constraints that symbolically 
represent \emph{all} possible lists up to a given depth, 
using propositional \emph{choice variables} to 
symbolically pick between these two alternatives.
Every (satisfying) assignment of choices returned by 
the SMT solver gives \toolname the concrete data and 
constructors used at each level, allowing it to decode 
the assignment into a Haskell value.

For example, \toolname represents valid @[(Pos, Score)]@ 
inputs (of depth up to 3), required to test @average@, 
as the conjunction of $\cstr{C_{list}}$ and $\cstr{C_{data}}$:
\begin{eqnarray*}
\cstr{C_{list}} & \defeq & (\cvar{c}_{00} \Rightarrow \cvar{xs}_0 = \lnil) \wedge 
                          (\cvar{c}_{01} \Rightarrow \cvar{xs}_0 = \lcons{\cvar{x}_1}{\cvar{xs}_1}) \wedge 
                          (\cvar{c}_{00} \oplus \cvar{c}_{01}) \\
               & \wedge & (\cvar{c}_{10} \Rightarrow \cvar{xs}_1 = \lnil) \wedge
                          (\cvar{c}_{11} \Rightarrow \cvar{xs}_1 = \lcons{\cvar{x}_2}{\cvar{xs}_2}) \wedge 
                          (\cvar{c}_{01} \Rightarrow \cvar{c}_{10} \oplus \cvar{c}_{11}) \\
               & \wedge & (\cvar{c}_{20} \Rightarrow \cvar{xs}_2 = \lnil) \wedge 
                          (\cvar{c}_{21} \Rightarrow \cvar{xs}_2 = \lcons{\cvar{x}_3}{\cvar{xs}_3}) \wedge 
                          (\cvar{c}_{11} \Rightarrow \cvar{c}_{20} \oplus \cvar{c}_{21}) \\
               & \wedge & (\cvar{c}_{30} \Rightarrow \cvar{xs}_3 = \lnil) \wedge 
                          (\cvar{c}_{21} \Rightarrow \cvar{c}_{30}) \\[0.1in]
\cstr{C_{data}} & \defeq & (\cvar{c}_{01} \Rightarrow \cvar{x}_1 = \ltup{\cvar{w}_1}{\cvar{s}_1} \ \wedge\ 0 < \cvar{w}_1 \ \wedge\ 0 \leq \cvar{s}_1 < 100) \\
               & \wedge & (\cvar{c}_{11} \Rightarrow \cvar{x}_2 = \ltup{\cvar{w}_2}{\cvar{s}_2} \ \wedge\ 0 < \cvar{w}_2 \ \wedge\ 0 \leq \cvar{s}_2 < 100) \\
               & \wedge & (\cvar{c}_{21} \Rightarrow \cvar{x}_3 = \ltup{\cvar{w}_3}{\cvar{s}_3} \ \wedge\ 0 < \cvar{w}_3 \ \wedge\ 0 \leq \cvar{s}_3 < 100)
\end{eqnarray*}
The first set of constraints $\cstr{C_{list}}$ describes all lists up to 
size 3. At each level $i$, the \emph{choice} variables $\cvar{c}_{i0}$ 
and $\cvar{c}_{i1}$ determine whether at that level the constructed 
list $\cvar{xs}_i$ is a ``nil'' or a ``cons''. 
In the constraints $\lnil$ and $(\lcons{}{})$ are \emph{uninterpreted} 
functions that represent ``nil'' and ``cons'' respectively. 
These functions only obey the congruence axiom and hence, can be 
efficiently analyzed by SMT solvers~\cite{Nelson81}.
The data at each level $\cvar{x}_i$ is constrained to be a pair of a 
positive weight $\cvar{w}_i$ and a valid score $\cvar{s}_i$.

The choice variables at each level are used to \emph{guard} the 
constraints on the next levels. 
First, if we are generating a ``cons'' at a given level, then
exactly one of the choice variables for the next level must be 
selected;
\eg  $\cvar{c}_{11} \Rightarrow \cvar{c}_{20} \oplus \cvar{c}_{21}$.
Second, the constraints on the data at a given level only hold 
if we are generating values for that level; \eg $\cvar{c}_{21}$ 
is used to guard the constraints on $\cvar{x}_3$, $\cvar{w}_3$ 
and $\cvar{s}_3$.
This is essential to avoid over-constraining the system 
which would cause \toolname to miss certain tests.

To \emph{decode} a model of the above into a Haskell value of type @[(Int, Int)]@,
we traverse constraints and use the valuations of the choice variables to 
build up the list appropriately.
At each level, if $\cvar{c}_{i0} \mapsto \ttrue$, then the list at that 
level is @[]@, otherwise $\cvar{c}_{i1} \mapsto \ttrue$ and we decode 
$\cvar{x}_{i+1}$ and $\cvar{xs}_{i+1}$ and ``cons'' the results.

We can iteratively generate \emph{multiple} inputs by adding a constraint that
refutes each prior model. As an important optimization, we only refute the
relevant parts of the model, \ie those needed to construct the list
(\S~\ref{sec:refute}).

\mypara{Ordered Containers}
Next, let us see how \toolname enables automatic testing with 
highly constrained inputs, such as the \emph{increasingly ordered} 
@OrdList@ values required by @insert@.
From the type definition, it is apparent that ordered
lists are the same as the usual lists described by
$\cstr{C_{list}}$, except that each unfolded \emph{tail} 
must only contain values that are greater than the 
corresponding \emph{head}.
That is, as we unfold @x1:x2:xs :: OrdList@ 
\begin{itemize}
\item At level @0@, we have @OrdList {v:Score| true}@
\item At level @1@, we have @OrdList {v:Score| x1 <= v}@
\item At level @2@, we have @OrdList {v:Score| x2 <= v && x1 <= v}@
\end{itemize}
and so on. Thus, we encode @OrdList Score@ (of depth up to 3) by
conjoining $\cstr{C_{list}}$ with  $\cstr{C_{score}}$ and $\cstr{C_{ord}}$,
which capture the valid score and ordering requirements respectively:
\begin{eqnarray*}
\cstr{C_{ord}}   & \defeq & (\cvar{c}_{11} \Rightarrow \cvar{x}_1 \leq \cvar{x}_2)
                \ \wedge\  (\cvar{c}_{21} \Rightarrow \cvar{x}_2 \leq \cvar{x}_3\ \wedge\ \cvar{x}_1 \leq \cvar{x}_3) \\[0.01in]
\cstr{C_{score}} & \defeq & (\cvar{c}_{01} \Rightarrow 0 \leq \cvar{x}_1 < 100)
                \ \wedge\  (\cvar{c}_{11} \Rightarrow 0 \leq \cvar{x}_2 < 100)
                \ \wedge\  (\cvar{c}_{21} \Rightarrow 0 \leq \cvar{x}_3 < 100)
\end{eqnarray*}

\mypara{Structured Containers}
Recall that @best k@ requires inputs whose \emph{structure} is constrained -- the 
size of the list should be no less than @k@. We specify size using special measure 
functions~\cite{VazouICFP14}, which let us relate the size of a list with that of
its unfolding, and hence, let us encode the notion of size inside the constraints:
\begin{eqnarray*}
\cstr{C_{size}} & \defeq & (\cvar{c}_{00} \Rightarrow \clen{\cvar{xs}_{0}} = 0) \wedge 
                          (\cvar{c}_{01} \Rightarrow \clen{\cvar{xs}_{0}} = 1 + \clen{\cvar{xs}_1}) \\
               & \wedge & (\cvar{c}_{10} \Rightarrow \clen{\cvar{xs}_{1}} = 0) \wedge 
                          (\cvar{c}_{11} \Rightarrow \clen{\cvar{xs}_{1}} = 1 + \clen{\cvar{xs}_2}) \\
               & \wedge & (\cvar{c}_{20} \Rightarrow \clen{\cvar{xs}_{2}} = 0) \wedge 
                          (\cvar{c}_{21} \Rightarrow \clen{\cvar{xs}_{2}} = 1 + \clen{\cvar{xs}_3}) \\
               & \wedge & (\cvar{c}_{30} \Rightarrow \clen{\cvar{xs}_{3}} = 0)
\end{eqnarray*}
At each unfolding, we instantiate the definition of the measure 
for each alternative of the datatype. 
In the constraints, $\clen{\cdot}$ is an uninterpreted function derived
from the measure definition. All of the relevant properties of the function
are spelled out by the unfolded constraints in $\cstr{C_{size}}$ and hence,
we can use SMT to search for models for the above constraint.
Hence, \toolname constrains the input type for @best@ as:
$$     0 \leq k 
\wedge \cstr{C_{list}} 
\wedge \cstr{C_{score}} 
\wedge \cstr{C_{size}} 
\wedge k \leq \clen{\cvar{xs}_0} $$
where the final conjunct comes from the top-level refinement that 
stipulates the input have at least @k@ scores.
Thus, \toolname only generates lists that are large enough. 
For example, in any model where $k = 2$, it will \emph{not} 
generate the empty or singleton list, as in those cases, 
$\clen{\cvar{xs}_0}$ would be $0$ (resp. $1$), violating the 
final conjunct above.

\mypara{Higher-order Functions}
Finally, \toolname's type-directed testing scales up to higher-order
functions using the same insight as in QuickCheck~\cite{claessen_quickcheck:_2000}, namely, 
to generate a function it suffices to be able to 
generate the \emph{output} of the function.
When tasked with the generation of a functional argument @f@, \toolname 
returns a Haskell function that when executed checks
whether its inputs satisfy @f@'s pre-conditions.
If they do, then @f@ uses \toolname to dynamically
query the SMT solver for an output that satisfies the 
constraints imposed by the concrete inputs.
Otherwise, @f@'s specifications are violated
and TARGET reports a counterexample.

This concludes our high-level tour of the benefits and 
implementation of \toolname. 
Notice that the property specification mechanism -- 
refinement types -- allowed us to get immediate feedback
that helped debug not just the code, but also the specification 
itself. 
Additionally, the specifications gave us machine-readable 
documentation about the behavior of functions, and a large 
unit test suite with which to automatically validate the 
implementation.
Finally, though we do not focus on it here, the specifications 
are amenable to formal verification should the programmer 
so desire.


\section{A Framework for Type Targeted Testing}\label{sec:framework}

Next, we describe a framework for type targeted testing, by formalizing
an abstract representation of refinement types~(\S~\ref{sec:reftypes}), 
describing the operations needed to generate tests from types~(\S~\ref{sec:targetable}), 
and then using the above to implement \toolname via a query-decode-check 
loop~(\S~\ref{sec:loop}). 
Subsequently, we instantiate the framework to obtain tests
for refined primitive types, lists, algebraic datatypes and higher-order 
functions~(\S~\ref{sec:list}).

 
\subsection{Refinement Types}\label{sec:reftypes}

\begin{figure}[t!]
\begin{mdframed}
\begin{CenteredBox}
\begin{code}
-- Manipulating Refinements
refinement :: RefType -> Refinement
subst      :: RefType -> [(Var, Var)] -> RefType

-- Manipulating Types
unfold     :: Ctor  -> RefType -> [(Var, RefType)]
binder     :: RefType -> Var
proxy      :: RefType -> Proxy a 
\end{code}
\end{CenteredBox}
\end{mdframed}
\caption{Refinement Type API}\label{fig:rtype}
\end{figure}

A refinement type is a type, where each component is 
decorated with a predicate from a refinement logic. 
For clarity, we describe refinement types and refinements 
abstractly as @RefType@ and @Refinement@ respectively.
We write @Var@ as an alias for @Refinement@ that is 
typically used to represent logical variables appearing
within the refinement.

\mypara{Notation} 
In the sequel, we will use 
double brackets $\meta{}$ to represent the 
various entities in the meta-language used to describe \toolname. 
For example, 
|$\meta{k}$|,
|$\meta{k \leq len\ v}$|, and
|$\meta{\reft{v}{[Score]}{k \leq len\ v}}$|
are the @Var@, @Refinement@, and @RefType@
representing the corresponding entities written in the brackets.

Next, we describe the various operations over them 
needed to implement \toolname.
These operations, summarized in Figure~\ref{fig:rtype}, 
fall into two categories: those which manipulate the 
\emph{refinements} and those which manipulate the 
\emph{types}.

\mypara{Operating on Refinements} 
To generate constraints and check inhabitation, we use 
the function @refinement@ which returns the (top-level) refinement
that decorates the given refinement type.
We will generate fresh @Var@s to name values of components, and will 
use @subst@ to replace the free occurrences of variables in a given \hbox{@RefType@.}
Suppose that @t@ is the @RefType@ represented by
\hbox{|$\meta{\reft{v}{[Score]}{k \leq len\ v}}$|.} Then,
\begin{itemize}
\item{@refinement t@} evaluates to |$\meta{k \leq len\ v}$| and
\item{|subst t [($\meta{k}$, $\meta{x_0}$)]|} evaluates to |$\meta{\reft{v}{[Score]}{x_0 \leq len\ v}}$|.
\end{itemize}

\mypara{Operating on Types} 
To build up compound values (\eg lists) from components 
(\eg an integer and a list), 
@unfold@ breaks a @RefType@ (\eg a list of integers) into its 
constituents (\eg an integer and a list of integers) at a given 
constructor (\eg ``cons'').
@binder@ simply extracts the @Var@ representing the
value being refined from the \hbox{@RefType@.}
To write generic functions over @RefType@s and use Haskell's
type class machinery to @query@ and @decode@ components of
types, we associate with each refinement type a \emph{proxy}
representing the corresponding Haskell type (in practice
this must be passed around as a separate argument).
For example, if @t@ is \hbox{|$\meta{\reft{v}{[Score]}{k \leq len\ v}}$|,} 
\begin{itemize}
\item{|unfold $\meta{:}$ t|} evaluates to |[($\meta{x}$, $\meta{Score}$), ($\meta{xs}$, $\meta{[Score]}$)]|,
\item{@binder t@} evaluates to |$\meta{v}$|, and
\item{@proxy t@} evaluates to a value of type @Proxy [Int]@.
\end{itemize}

\subsection{The \texttt{Targetable} Type Class}\label{sec:targetable}


Following \quickcheck, we encapsulate the key operations needed
for type-targeted testing in a type class @Targetable@ 
(Figure~\ref{fig:targetable}). 
\begin{figure}
\begin{mdframed}
\begin{CenteredBox}
\begin{code}
class Targetable a where
  query  :: Proxy a -> Int -> RefType -> SMT Var
  decode :: Var -> SMT a
  check  :: a -> RefType -> SMT (Bool, Var)
  toReft :: a -> Refinement
\end{code}
\end{CenteredBox}
\end{mdframed}
\caption{The class of types that can be tested by \toolname}\label{fig:targetable}
\end{figure}
This class characterizes the set of types which can be tested 
by \toolname. All of the operations can interact with an external SMT 
solver, and so return values in an @SMT@ monad.

\begin{itemize}
\item{@query@} takes a \emph{proxy} for the Haskell type
   for which we are generating values, an integer 
   \emph{depth} bound, and a \emph{refinement type}
   describing the desired constraints, and generates a set of 
   logical constraints and a @Var@ that represents the 
   constrained value.

\item{@decode@} takes a @Var@, generated via a previous 
   @query@ and queries the model returned by the SMT solver
   to construct a Haskell value of type @a@.
 
\item{@check@} takes a value of type @a@, translates 
   it back into logical form, and verifies that it inhabits
   the output type @t@.
   
\item{@toReft@} takes a value of type @a@ and translates it
   back into logical form (a specialization of @check@).
\end{itemize}

\subsection{The Query-Decode-Check Loop}\label{sec:loop}

Figure~\ref{fig:arch} summarizes the overall implementation of 
\toolname, which takes as input a function @f@ and its refinement 
type specification @t@ and proceeds to test the function against 
the specification via a \emph{query-decode-check} loop:
(1) First, we translate the refined @inputTypes@ into a logical \emph{query}.
(2) Next, we \emph{decode} the model (\ie satisfying assignment) for the 
    query returned by the SMT solver to obtain concrete @inputs@.
(3) Finally, we @execute@ the function @f@ on the @inputs@ to get the 
    corresponding @output@, which we @check@ belongs to the specified 
    @outputType@. If the @check@ fails, we return the @inputs@ as a counterexample.
After each test, \toolname, refutes the given test to force the SMT 
solver to return a different set of inputs, and this process is repeated until 
a user specified number of iterations. The @checkSMT@ call may fail
to find a model meaning that we have exhaustively tested all inputs upto
a given @testDepth@ bound. If all iterations succeed, \ie no counterexamples
are found, then \toolname returns @Ok@, indicating that @f@ satisfies @t@ 
up to the given depth bound.

\begin{figure}[ht!]
\begin{mdframed}
\begin{CenteredBox}
\begin{code} 
target f t = do 
  let txs = inputTypes t
  vars  <- forM txs $ \tx -> 
             query (proxy tx) testDepth tx -- Query
  forM [1 .. testNum] $ \_ -> do
    hasModel <- checkSMT 
    when hasModel $ do
      inputs <- forM vars decode           -- Decode
      output <- execute f inputs                        
      let su = zip (map binder txs) (map toReft inputs)
      let to = outputType t `subst` su
      (ok,_) <- check output to            -- Check
      if ok then 
        refuteSMT 
      else 
        throw (CounterExample inputs)     
  return Ok
\end{code}
\end{CenteredBox}
\end{mdframed}
\caption{Implementing \toolname via a \emph{query-decode-check} loop}\label{fig:arch}
\end{figure}

\section{Instantiating the \toolname Framework}\label{sec:list}

Next, we describe a concrete instantiation of \toolname for lists.
We start with a constraint generation API~(\S~\ref{sec:constraint}). 
Then we use the API to implement the key operations 
\hbox{@query@~(\S~\ref{sec:query}),} 
\hbox{@decode@~(\S~\ref{sec:decode}),} 
\hbox{@check@~(\S~\ref{sec:check}),} and
\hbox{@refuteSMT@~(\S~\ref{sec:refute}),} 
thereby enabling \toolname to automatically test functions over lists.
We omit the definition of @toReft@ as it follows directly from the
definition of @check@.
Finally, we show how the list instance can be generalized to algebraic 
datatypes and higher-order functions~(\S~\ref{sec:generic}).

\subsection{SMT Solver Interface}\label{sec:constraint}

Figure~\ref{fig:smt} describes the interface to the SMT 
solvers that \toolname uses for constraint generation and 
model decoding. The interface has functions to 
(a)~generate logical variables of type @Var@, 
(b)~constrain their values using @Refinement@ predicates, and
(c)~determine the values assigned to the variables in satisfying models.

\begin{figure}[ht!]
\begin{mdframed}
\begin{CenteredBox}
\begin{code} 
fresh     :: SMT Var
guard     :: Var -> SMT a      -> SMT a 
constrain :: Var -> Refinement -> SMT ()

apply     :: Ctor -> [Var] -> SMT Var 
unapply   :: Var  -> SMT (Ctor, [Var])

oneOf     :: Var -> [(Var, Var)] -> SMT ()
whichOf   :: Var -> SMT Var

eval      :: Refinement -> SMT Bool
\end{code}
\end{CenteredBox}
\end{mdframed}
\caption{SMT Solver API}\label{fig:smt}
\end{figure}

\begin{itemize}

\item{@fresh@} allocates a new logical variable.

\item{@guard b act@} ensures that all the constraints 
generated by @act@ are \emph{guarded by} the choice 
variable @b@. That is, if @act@ generates the constraint 
$p$ then @guard b act@ generates the (implication) 
constraint ${b \Rightarrow p}$.

\item{@constrain x r@} generates a constraint that @x@ 
satisfies the refinement predicate @r@.

\item{@apply c xs@} generates a new @Var@ for the folded up value 
obtained by applying the constructor @c@ to the fields @xs@,
while also generating constraints from the measures. For example, 
{|apply $\meta{:}$ [$\meta{x_1}$, $\meta{xs_1}$]|} returns |$\meta{\lcons{\cvar{x}_1}{\cvar{xs}_1}}$|
and generates the constraint
${\clen{(\lcons{\cvar{x}_1}{\cvar{xs}_1})} = 1 + \clen{\cvar{xs}_1}}$.

\item{@unapply x@} returns the @Ctor@ and @Var@s from which the input 
@x@ was constructed. 

\item{@oneOf x cxs@} generates a constraint that @x@ equals exactly
one of the elements of @cxs@. For example, 
{|oneOf $\meta{xs_0}$ [($\meta{c_{00}}$,$\meta{[]}$),($\meta{c_{01}}$,$\meta{x_1 : xs_1}$)]|} 
yields:
$$(\cvar{c}_{00} \Rightarrow \cvar{xs}_0 = \lnil) \wedge 
  (\cvar{c}_{01} \Rightarrow \cvar{xs}_0 = \lcons{\cvar{x}_1}{\cvar{xs}_1}) \wedge 
  (\cvar{c}_{00} \oplus \cvar{c}_{01})$$

\item{@whichOf x@} returns the particular alternative that was 
assigned to @x@ in the current model returned by the 
SMT solver. Continuing the previous example, if the model sets 
|$\meta{c_{00}}$| (resp. |$\meta{c_{01}}$|) to $\ttrue$, |whichOf $\meta{xs_0}$| returns 
|$\meta{[]}$| (resp \hbox{|$\meta{x_1 : xs_1}$|).}

\item{@eval r@} checks the validity of a refinement with no free variables. For
  example, |eval $\meta{len\ (1 : []) > 0}$| would return @True@.

\end{itemize}

\subsection{Query}\label{sec:query}

Figure~\ref{fig:query} shows the procedure for constructing a 
@query@ from a refined list type, \eg the one required as an input 
to the @best@ or @insert@ functions from \S~\ref{sec:overview}.

\begin{figure}[t!]
\begin{mdframed}
\begin{CenteredBox}
\begin{mcode}
query p d t = do
  let cs = ctors d
  bs <- forM cs (\_ -> fresh)
  xs <- zipWithM (queryCtor (d-1) t) bs cs
  x  <- fresh 
  oneOf x     (zip bs xs)
  constrain x (refinement t)
  return x

queryCtor d t b c = guard b (do
  let fts = unfold c t
  fs'    <- scanM (queryField d) [] fts
  x      <- apply c fs'
  return x)

queryField d su (f, t) = do
  f' <- query (proxy t) d (t `subst` su)
  return ((f, f') : su, f')                    
ctors d
  | d > 0     = [ $\meta{:}$, $\meta{[]}$ ]
  | otherwise = [ $\meta{[]}$ ]
\end{mcode}
\end{CenteredBox}
\end{mdframed}
\caption{Generating a Query}\label{fig:query}
\end{figure}

\mypara{Lists}
@query@ returns a @Var@ that represent \emph{all} lists up to 
depth @d@ that satisfy the logical constraints associated with 
the refined list type @t@.
To this end, it invokes @ctors@ to obtain all of the suitable
constructors for depth @d@. For lists, when
the depth is @0@ we should only use the |$\meta{[]}$| constructor,
otherwise we can use either |$\meta{:}$| or |$\meta{[]}$|. 
This ensures that @query@ terminates after encoding all possible
lists up to a given depth \hbox{@d@.}
Next, it uses @fresh@ to generate a distinct \emph{choice} 
variable for each constructor, and calls \hbox{@queryCtor@ to}
generate constraints and a corresponding symbolic @Var@ 
for each constructor. 
The choice variable for each constructor is supplied to 
@queryCtor@ to ensure that the constraints are \emph{guarded}, 
\ie only required to hold \emph{if} the corresponding choice 
variable is selected in the model and not otherwise.
Finally, a fresh @x@ represents the value at depth @d@ and 
is constrained to be @oneOf@ the alternatives represented 
by the constructors, and to satisfy the top-level refinement of @t@.
%

\mypara{Constructors}
@queryCtor@ takes as input the refined list type @t@, 
a depth @d@, a particular constructor @c@ for the list 
type, and generates a query describing the \emph{unfolding}
of @t@ at the constructor @c@, guarded by the choice 
variable @b@ that determines whether this alternative 
is indeed part of the value.
These constraints are the conjunction of
those describing the values of the individual fields 
which can be combined via @c@ to obtain a @t@ value.
To do so, @queryCtor@ first @unfold@s the type @t@ at 
@c@, obtaining a list of constituent fields and their
respective refinement types @fts@. Next, it uses 
\begin{code}
  scanM :: Monad m => (a -> b -> m (a, c)) -> a -> [b] -> m [c]
\end{code}
to traverse the fields from left to right, building up 
representations of values for the fields from their 
unfolded refinement types.
Finally, we invoke @apply@ on @c@ and the fields @fs'@ to 
return a symbolic representation of the constructed value 
that is constrained to satisfy the measure properties of @c@.

\mypara{Fields}
@queryField@ generates the actual constraints for a
single field @f@ with refinement type @t@, by invoking
@query@ on @t@.  
The @proxy@ enables us to resolve the appropriate 
type-class instance for generating the query for 
the field's value.
Each field is described by a new symbolic name @f'@ which is 
@subst@ituted for the formal name of the field @f@ in the
refinements of subsequent fields, thereby tracking dependencies
between the fields.
For example, these substitutions ensure the values in 
the tail are greater than the head as needed by 
@OrdList@ from \S~\ref{sec:overview}.

\subsection{Decode}\label{sec:decode}
\begin{figure}[t!]
\begin{mdframed}
\begin{minipage}{0.45\textwidth}
\begin{CenteredBox}
\begin{code}
decode x = do 
  x'      <- whichOf x
  (c,fs') <- unapply x'
  decodeCtor c fs'
\end{code}
\end{CenteredBox}
\end{minipage}
\begin{minipage}{0.55\textwidth}
\begin{CenteredBox}
\begin{mcode}
decodeCtor $\meta{[]}$ []    = return []
decodeCtor $\meta{:}$ [x,xs] = do
  v  <- decode x
  vs <- decode xs
  return (v:vs)
\end{mcode}
\end{CenteredBox}
\end{minipage}
\end{mdframed}
\caption{Decoding Models into Haskell Values}\label{fig:decode}
\end{figure}
Once we have generated the constraints we query the SMT solver 
for a model, and if one is found we must \emph{decode} it into
a concrete Haskell value with which to test the given function.
Figure~\ref{fig:decode} shows how to decode an SMT model for lists. 

\mypara{Lists} @decode@ takes as input the top-level symbolic
representation @x@ and queries the model to determine which
alternative was assigned by the solver to @x@, \ie a nil or a cons.
Once the alternative is determined, we use @unapply@ to destruct 
it into its constructor @c@ and fields @fs'@, which are recursively
decoded by @decodeCtor@.

\mypara{Constructors} @decodeCtor@ takes the constructor @c@ and
a list of symbolic representations for fields, and decodes each 
field into a value and applies the constructor to obtain the 
Haskell value.
For example, in the case of the |$\meta{[]}$| constructor, there are no
fields, so we return the empty list. In the case of the |$\meta{:}$| 
constructor, we decode the head and the tail, and cons them to 
return the decoded value. 
@decodeCtor@ has the type
\begin{code}  
  Targetable a => Ctor -> [Var] -> SMT [a]
\end{code}
\ie if @a@ is a decodable type, then @decodeCtor@ suffices to decode lists of @a@.
Primitives like integers that are directly encoded 
in the refinement logic are the base case -- \ie the 
value in the model is directly translated into the 
corresponding Haskell value.

\subsection{Check}\label{sec:encode}\label{sec:check}
\begin{figure}[t!]
\begin{mdframed}
\begin{CenteredBox}
\begin{mcode}
check v t = do
  let (c,vs) = splitCtor v
  let fts    = unfold c t
  (bs, vs') <- fmap unzip (scanM checkField [] (zip vs fts))
  v'        <- apply c vs'
  let t'     = t `subst` [(binder t, v')]
  b'        <- eval (refinement t')
  return (and (b:bs), v')
  
checkField su (v, (f, t)) = do
  (b, v') <- check v (t `subst` su)
  return ((f, v') : su, (b, v'))

splitCtor []     = ($\meta{[]}$, [])
splitCtor (x:xs) = ($\meta{:}$, [x,xs])
\end{mcode}
\end{CenteredBox}
\end{mdframed}
\caption{Checking Outputs}\label{fig:check}
\end{figure}
The third step of the query-decode-check loop is to verify
that the output produced by the function under test indeed
satisfies the output refinement type of the function.
We accomplish this by \emph{encoding} the output value as a
logical expression, and evaluating the output refinement
applied to the logical representation of the output value.

@check@, shown in Figure~\ref{fig:check}, takes a Haskell
(output) value @v@ and the (output) refinement type @t@, and
recursively verifies each component of the output type. It
converts each component into a logical representation,
@subst@itutes the logical expression for the symbolic value,
and @eval@uates the resulting @Refinement@.



\subsection{Refuting Models} \label{sec:refute}

Finally, \toolname invokes @refuteSMT@ to \emph{refute} a 
given model in order to force the SMT solver to produce a 
different model that will yield a different test input.
A na\"{\i}ve implementation of refutation is as follows.
Let $X$ be the set of all variables appearing in the constraints.
Suppose that in the current model, each variable $x$ is assigned 
the value $\val{x}$.
Then, to refute the model, we add a \emph{refutation constraint} 
$
\vee_{x \in X} x \not = \val{x}
$.
That is, we stipulate that \emph{some} variable be assigned a 
different value.

The na\"{\i}ve  implementation is extremely inefficient.
The SMT solver is free to pick a different value for some 
\emph{irrelevant} variable which was not even used for decoding.
As a result, the next model can, after decoding, yield the 
\emph{same} Haskell value, thereby blowing up the number of 
iterations needed to generate all tests of a given size.

\toolname solves this problem by forcing the SMT solver to return
models that yield \emph{different decoded tests} in each iteration.
To this end \toolname restricts the refutation constraint to 
the set of variables that were actually used to @decode@ the 
Haskell value.
We track this set by instrumenting the @SMT@ monad to log the 
set of variables and choice-variables that are transitively 
queried via the recursive calls to @decode@.
That is, each call to @decode@ logs its argument, and each call 
to @whichOf@ logs the choice variable corresponding to the 
alternative that was returned.
Let $R$ be the resulting set of \emph{decode-relevant} variables.
\toolname refutes the model by using a \emph{relevant refutation constraint}
$
\vee_{x \in R} x \not = \val{x}
$
which ensures that the next model decodes to a different value.

\subsection{Generalizing \toolname To Other Types}\label{sec:generic}

The implementation in \S~\ref{sec:list} is for 
List types, but @ctors@, @decodeCtor@, and @splitCtor@ are the only 
functions that are List-specific. 
Thus, we can easily generalize the implementation to:
\begin{itemize}
\item{\emph{primitive datatypes}}, \eg integers, by returning an empty 
    list of constructors,
\item{\emph{algebraic datatypes}}, by implementing @ctors@, @decodeCtor@, and @splitCtor@ for that type.
\item{\emph{higher-order functions}}, by lifting instances of @a@ to functions returning @a@.
\end{itemize}

\mypara{Algebraic Datatypes}
Our List implementation has three pieces of type-specific logic:
\begin{itemize}
\item{@ctors@}, which returns a list of constructors to unfold;
\item{@decodeCtor@}, which decodes a specific @Ctor@; and
\item{@splitCtor@}, which splits a Haskell value into a pair of its @Ctor@ and fields. 
\end{itemize}

Thus, to instantiate \toolname on a new data type, all we need is to 
implement these three operations for the type. This implementation
essentially follows the concrete template for Lists.
In fact, we observe that the recipe is entirely mechanical boilerplate,  
and can be fully automated for \emph{all} algebraic data types by using 
a \emph{generics} library.

Any algebraic datatype (ADT) can be represented as a \emph{sum-of-products} 
of component types. A generics library, such as \GhcGenerics~\cite{magalhaes_generic_2010}, 
provides a \emph{univeral} sum-of-products type and functions to automatically 
convert any ADT to and from the universal representation.
Thus, to obtain @Targetable@ instances for \emph{any} ADT it suffices
to define a @Targetable@ instance for the \emph{universal} type.

Once the universal type is @Targetable@ we can automatically get an 
instance for any new user-defined ADT (that is an instance of @Generic@) as follows:
(1)~to generate a \emph{query} we simply create a query for 
    \GhcGenerics' universal representation of the refined type,
(2)~to \emph{decode} the results from the SMT solver, we 
    decode them into the universal representation and then use 
    \GhcGenerics to map them back into the user-defined type,
(3)~to \emph{check} that a given value inhabits a user-defined 
    refinement type, we check that the universal representation 
    of the value inhabits the type's universal counterpart.

The @Targetable@ instance for the universal representation is a 
generalized version of the List instance from \S~\ref{sec:list}, 
that relies on various technical details of \GhcGenerics.

\mypara{Higher Order Functions} 
Our type-directed approach to specification makes it easy to extend
\toolname to higher-order functions. Concretely, it suffices to 
implement a type-class instance:
\begin{code}
  instance (Targetable input, Targetable output) 
    => Targetable (input -> output)
\end{code}
In essence, this instance uses the @Targetable@ 
instances for @input@ and @output@ to 
create an instance for functions from @input -> output@,
after which Haskell's type class machinery suffices to 
generate concrete function values.

To create such instances, we use the insight from 
\quickcheck, that to generate (constrained) functions,
we need only to generate \emph{output} values for the function. 
Following this route, we generate functions by creating 
new lambdas that take in the inputs from the calling context, 
and use their values to create queries for the output, after 
which we can call the SMT solver and decode the results 
to get concrete outputs that are returned by the lambda, 
completing the function definition. 
Note that we require @input@ to also be @Targetable@
so that we can encode the Haskell value in the refinement logic,
in order to constrain the output values suitably.
We additionally memoize the generated function to preserve the 
illusion of purity. 
It is also possible to, in the future, extend our 
implementation to refute functions by asserting 
that the output value for a given input be distinct 
from any previous outputs for that input.

\section{Evaluation} \label{sec:evaluation}

We have built a prototype implementation of \toolname\footnote{\url{http://hackage.haskell.org/package/target-0.1.1.0}} and next, 
describe an evaluation on a series of benchmarks ranging from 
textbook examples of algorithms and data structures to widely 
used Haskell libraries like \textsc{containers} and \textsc{xmonad}.
Our goal in this evaluation is two-fold. 
First, we describe micro-benchmarks (\ie functions)
that \emph{quantitatively compare} \toolname with 
the existing state-of-the-art, property-based testing
tools for Haskell -- namely \smallcheck and \quickcheck\ -- 
to determine whether \toolname is indeed able to generate
highly constrained inputs more effectively.
Second, we describe macro-benchmarks (\ie modules) that 
evaluate the amount of \emph{code coverage} that we 
get from type-targeted testing.
%


\begin{figure}[ht!]
  \centering

  \begin{tikzpicture}
    \begin{groupplot}[
      group style = {group size = 3 by 1, horizontal sep=15pt,},
      groupplot ylabel={Time (sec)},
      groupplot xlabel={Depth},
      group/only outer labels,
      ymode=log,
      ymax=10000,
      ymin=0.0001
    ]
    \nextgroupplot[
      title=\textsc{List.insert}
    ]
    \addplot table[smooth,col sep=comma,x index=0,y index=1] {csv/List.insert.csv};
    \addplot table[smooth,col sep=comma,x index=0,y index=2] {csv/List.insert.csv};
    \addplot table[smooth,col sep=comma,x index=0,y index=3] {csv/List.insert.csv};
    \nextgroupplot[
      title=\textsc{RBTree.add},
      legend columns=4,
      legend entries={\toolname,\smallcheck,\lazysmallcheck,\lazysmallcheck (slow)},
      legend to name=legend,
    ]
    \addplot table[smooth,col sep=comma,x index=0,y index=1] {csv/RBTree.add.csv};
    \addplot table[smooth,col sep=comma,x index=0,y index=2] {csv/RBTree.add.csv};
    \addplot table[smooth,col sep=comma,x index=0,y index=3] {csv/RBTree.add.csv};
    \addplot table[smooth,col sep=comma,x index=0,y index=4] {csv/RBTree.add.csv};
    \nextgroupplot[
      title=\textsc{XMonad.focus\_left}
    ]
    \addplot table[smooth,col sep=comma,x index=0,y index=1] {csv/XMonad.focus_left.csv};
    \addplot table[smooth,col sep=comma,x index=0,y index=2] {csv/XMonad.focus_left.csv};
    \addplot table[smooth,col sep=comma,x index=0,y index=3] {csv/XMonad.focus_left.csv};
    \end{groupplot}
  \end{tikzpicture}
  \begin{tikzpicture}
    \begin{groupplot}[
      group style = {group size = 2 by 1, horizontal sep=15pt,},
      groupplot ylabel={Time (sec)},
      groupplot xlabel={Depth},
      group/only outer labels,
      ymode=log,
      ymax=10000,
      ymin=0.0001
    ]
    \nextgroupplot[
      title=\textsc{Map.delete}
    ]
    \addplot table[smooth,col sep=comma,x index=0,y index=1] {csv/Map.delete.csv};
    \addplot table[smooth,col sep=comma,x index=0,y index=2] {csv/Map.delete.csv};
    \addplot table[smooth,col sep=comma,x index=0,y index=3] {csv/Map.delete.csv};
    \addplot table[smooth,col sep=comma,x index=0,y index=4] {csv/Map.delete.csv};
    \nextgroupplot[
      title=\textsc{Map.difference}
    ]
    \addplot table[smooth,col sep=comma,x index=0,y index=1] {csv/Map.difference.csv};
    \addplot table[smooth,col sep=comma,x index=0,y index=2] {csv/Map.difference.csv};
    \addplot table[smooth,col sep=comma,x index=0,y index=3] {csv/Map.difference.csv};
    \addplot table[smooth,col sep=comma,x index=0,y index=4] {csv/Map.difference.csv};
    \end{groupplot}
  \end{tikzpicture}\\
  \ref{legend}

  \caption{Results of comparing \toolname with \quickcheck, \smallcheck, and Lazy
    \smallcheck on a series of functions. \toolname, \smallcheck, and Lazy
    \smallcheck were both configured to check the first 1000 inputs that
    satisfied the precondition at increasing depth parameters, with a 60 minute
    timeout per depth; \quickcheck was run with the default settings, \ie it had
    to produce 100 test cases. \toolname, \smallcheck, and \lazysmallcheck were
    configured to use the same notion of depth, in order to ensure they would
    generate the same number of valid inputs at each depth level. \quickcheck was
    unable to successfully complete any run due to the low probability of
    generating valid inputs at random.}\label{fig:comparisonresults}
\end{figure}

\subsection{Comparison with \quickcheck and \smallcheck}\label{sec:comparison}

We compare \toolname with \quickcheck and \smallcheck by using 
a set of benchmarks with highly constrained inputs. 
For each benchmark we compared \toolname with \smallcheck and
\quickcheck, with the latter two using the generate-and-filter 
approach, wherein a value is generated and subsequently discarded if
it does not meet the desired constraint.
While one could possibly write custom ``operational'' generators 
for each property, the point of this evaluation is compare the 
different approaches ability to enable ``declarative'' specification 
driven testing.
Next, we describe the benchmarks and then summarize the results of the comparison
(Figure~\ref{fig:comparisonresults}).

\mypara{Inserting into a sorted \List}
Our first benchmark is the \Insert function from the homonymous 
sorting routine. We use the specification that given an element 
and a sorted list, @insert x xs@ should evaluate to a sorted list.
We express this with the type
\begin{code}
  type Sorted a = List <{\hd v -> hd < v}> a
  insert :: a -> Sorted a -> Sorted a
\end{code}
where the ordering constraint is captured by an abstract 
refinement~\cite{Vazou13} which states that \emph{each} 
list head @hd@ is less than every element @v@ in its tail.

\mypara{Inserting into a Red-Black Tree}
Next, we consider insertion into a Red-Black tree.
\begin{code}
  data RBT a = Leaf  | Node Col a (RBT a) (RBT a)
  data Col   = Black | Red
\end{code}
Red-black trees must satisfy three invariants:
(1)~red nodes always have black children,
(2)~the black height of all paths from the root to a leaf is the same, and
(3)~the elements in the tree should be ordered.
We capture (1) via a measure that recursively checks each @Red@ node has @Black@ children.
\begin{code}
  measure isRB :: RBT a -> Prop
  isRB Leaf           = true
  isRB (Node c x l r) = isRB l && isRB r &&
                        (c == Red => isBlack l && isBlack r)
\end{code}
We specify (2) by defining the @Black@ height as:
\begin{code}
  measure bh :: RBT a -> Int
  bh Leaf           = 0
  bh (Node c x l r) = bh l + (if c == Red then 0 else 1)
\end{code}
and then checking that the @Black@ height of both subtrees is the same:
\begin{code}
  measure isBH :: RBT a -> Prop
  isBH Leaf           = true
  isBH (Node c x l r) = isBH l && isBH r && bh l == bh r
\end{code}
Finally, we specify the (3), the ordering invariant as:
\begin{code}
  type OrdRBT a = RBT <{\r v -> v < r}, {\r v -> r < v}> a
\end{code}
\ie with two abstract refinements for the left and right subtrees
respectively, which state that the root @r@ is greater than (resp. less than)
each element @v@ in the subtrees. Finally, a valid Red-Black tree is:
\begin{code}
  type OkRBT a = {v:OrdRBT a | isRB v && isBH v}
\end{code}
Note that while the specification for the \emph{internal} invariants for Red-Black
trees is tricky, the specification for the public API -- \eg the @add@ function -- 
is straightforward:
\begin{code}
  add :: a -> OkRBT a -> OkRBT a
\end{code}

\mypara{Deleting from a Data.Map}\label{sec:delete-from-map}
Our third benchmark is the @delete@ function from the \hbox{@Data.Map@} module in 
the Haskell standard libraries. The @Map@ structure is a balanced binary
search tree that implements purely functional key-value dictionaries:
\begin{code}
  data Map k a = Tip | Bin Int k a (Map k a) (Map k a)
\end{code}
A valid @Data.Map@ must satisfy two properties:
(1)~the size of the left and right sub-trees must be 
    within a factor of three of each other, and
(2)~the keys must obey a binary search ordering.
We specify the balancedness invariant~(1) with a measure
\begin{code}
  measure isBal :: Map k a -> Prop
  isBal (Tip)           = true
  isBal (Bin s k v l r) = isBal l && isBal r &&
                          (sz l + sz r <= 1 ||
                           sz l <= 3 * sz r <= 3 * sz l)
\end{code}
and combine it with an ordering invariant (like @OrdRBT@) to specify valid trees.
\begin{code}
  type OkMap k a = {v : OrdMap k a | isBal v}
\end{code}
We can check that @delete@ preserves the invariants by 
checking that its output is an @OkMap k a@.
However, we can also go one step further and check 
the functional correctness property that @delete@ 
removes the given key, with a type:
\begin{code}
  delete :: Ord k => k:k -> m:OkMap k a 
         -> {v:OkMap k a | MinusKey v m k}
\end{code}
where the predicate @MinusKey@ is defined as:
\begin{code}
  predicate MinusKey M1 M2 K 
    = keys M1 = difference (keys M2) (singleton K)
\end{code}
using the measure @keys@ describing the contents of the @Map@:
\begin{code}
  measure keys :: Map k a -> Set k
  keys (Tip)           = empty () 
  keys (Bin s k v l r) = union (singleton k) 
                               (union (keys l) (keys r))
\end{code}

\mypara{Refocusing XMonad StackSets} \label{sec:refocus-stackset}
Our last benchmark comes from the tiling window manager XMonad. 
The key invariant of XMonad's internal @StackSet@ data structure 
is that the elements (windows) must all be \emph{unique}, \ie contain
no duplicates.
XMonad comes with a test-suite of over 100 \quickcheck properties;
we select one which states that moving the focus between windows 
in a @StackSet@ should not affect the \emph{order} of the windows.
\begin{code}
  prop_focus_left_master n s =
    index (foldr (const focusUp) s [1..n]) == index s
\end{code}
With \quickcheck, the user writes a custom generator for valid @StackSet@s
and then runs the above function on test inputs created by the generator, 
to check if in each case, the result of the above is @True@.

With \toolname, it is possible to test such properties \emph{without} 
requiring custom generators. Instead the user writes a declarative 
specification:
\begin{code}
  type OkStackSet = {v:StackSet | NoDuplicates v}
\end{code}
(We refer the reader to~\cite{VazouRealWorld14} for a full 
discussion of how to specify @NoDuplicates@).
Next, we define a refinement type:
\begin{code}
  type TTrue = {v:Bool | Prop v}
\end{code}
that is only inhabited by @True@, and use it to type the \quickcheck 
property as:
\begin{code}
  prop_focus_left_master :: Nat -> OkStackSet -> TTrue 
\end{code}
This property is particularly difficult to \emph{verify}; however,
\toolname is able to automatically
generate valid inputs to \emph{test} that @prop_focus_left_master@
always returns @True@.


\mypara{Results}
Figure~\ref{fig:comparisonresults} summarizes the results of the comparison.
\quickcheck was unable to successfully complete \emph{any} 
benchmark to the low probability of generating properly 
constrained values at random.

\begin{description}
\item[List Insert] \toolname is able to test @insert@ all the way to 
   depth 20, whereas \lazysmallcheck times out at depth 19.

\item[Red-Black Tree Insert] \toolname is able to test @add@ up to depth 12,
  while \lazysmallcheck times out at depth 6.
  
\item[Map Delete] \toolname is able to check @delete@ up to depth 10, whereas
   \lazysmallcheck times out at depth 7 if it checks ordering first,
    or depth 6 if it checks balancedness first.

\item[StackSet Refocus] \toolname and is able to check this property 
    up to depth 8, while \lazysmallcheck times out at depth 7.
\end{description}

\toolname sees a performance hit with properties 
that require reasoning with the theory of Sets \eg 
the no-duplicates invariant of @StackSet@. 
While \lazysmallcheck times out at a higher depths, when it completes
\eg at depth 6, it does so in 0.7s versus \toolname's 9 minutes.
We suspect this is because the theory of sets are a relatively recent
addition to SMT solvers \cite{arrayZ3}, and with further improvements 
in SMT technology, these numbers will get significantly better.

Overall, we found that for \emph{small inputs} \lazysmallcheck 
is substantially faster as exhaustive enumeration is tractable,
and does not incur the overhead of communicating with an external 
general-purpose solver.
Additionally, \lazysmallcheck benefits from pruning predicates 
that exploit laziness and only force a small portion of the 
structure (\eg ordering). 
However, we found that constraints that force the entire 
structure (\eg balancedness), or composing predicates in the 
wrong \emph{order}, can force \lazysmallcheck to enumerate 
the entire exponentially growing search space.

\toolname, on the other hand, scales nicely to larger input sizes,
allowing systematic and exhaustive testing of larger, more complex
inputs. This is because \toolname eschews \emph{explicit} 
enumeration-and-filtering (which results in searching for 
fewer needles in larger haystacks as the sizes increas), 
in favor of \emph{symbolically} searching for valid models 
via SMT, making \toolname robust to the strictness or ordering 
of constraints.

\subsection{Measuring Code Coverage}\label{sec:code-coverage}

The second question we seek to answer is whether \toolname is suitable for testing entire
libraries, \ie how much of the program can be automatically exercised using our
system? Keeping in mind the well-known issues with treating code coverage as an
indication of test-suite quality~\cite{marick1999misuse}, we
consider this experiment a negative filter.

To this end, we ran \toolname against the entire user-facing API of 
\hbox{@Data.Map@,} our @RBTree@ library, and @XMonad.StackSet@ -- using 
the constrained refined types (\eg @OkMap@, @OkRBT@, @OkStackSet@) as 
the specification for the exposed types -- and measured the expression 
and branch coverage, as reported by @hpc@~\cite{gill2007haskell}.
We used an increasing timeout ranging from one to thirty minutes
per exported function.

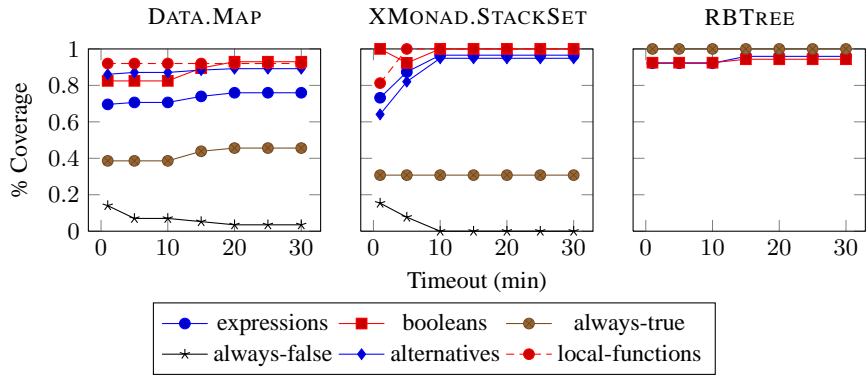
\begin{figure}[t!]
\centering
  \begin{tikzpicture}
    \begin{groupplot}[
      group style = {group size = 3 by 1, horizontal sep=15pt,},
      groupplot ylabel={\% Coverage},
      groupplot xlabel={Timeout (min)},
      group/only outer labels,
      ymin=0,
      ymax=1
    ]
    \nextgroupplot[
      title=\textsc{Data.Map},
      legend columns=3,
      legend entries={expressions,booleans,always-true,always-false,alternatives,local-functions},
      legend to name=legend,
    ]
    \addplot table[smooth,col sep=comma,x index=0,y index=1] {csv/MapCoverage.csv};
    \addplot table[smooth,col sep=comma,x index=0,y index=2] {csv/MapCoverage.csv};
    \addplot table[smooth,col sep=comma,x index=0,y index=3] {csv/MapCoverage.csv};
    \addplot table[smooth,col sep=comma,x index=0,y index=4] {csv/MapCoverage.csv};
    \addplot table[smooth,col sep=comma,x index=0,y index=5] {csv/MapCoverage.csv};
    \addplot table[smooth,col sep=comma,x index=0,y index=6] {csv/MapCoverage.csv};
    \nextgroupplot[
      title=\textsc{XMonad.StackSet},
    ]
    \addplot table[smooth,col sep=comma,x index=0,y index=1] {csv/StackSetCoverage.csv};
    \addplot table[smooth,col sep=comma,x index=0,y index=2] {csv/StackSetCoverage.csv};
    \addplot table[smooth,col sep=comma,x index=0,y index=3] {csv/StackSetCoverage.csv};
    \addplot table[smooth,col sep=comma,x index=0,y index=4] {csv/StackSetCoverage.csv};
    \addplot table[smooth,col sep=comma,x index=0,y index=5] {csv/StackSetCoverage.csv};
    \addplot table[smooth,col sep=comma,x index=0,y index=6] {csv/StackSetCoverage.csv};
    \nextgroupplot[
      title=\textsc{RBTree}
    ]
    \addplot table[smooth,col sep=comma,x index=0,y index=1] {csv/RBTreeCoverage.csv};
    \addplot table[smooth,col sep=comma,x index=0,y index=5] {csv/RBTreeCoverage.csv};
    \addplot table[smooth,col sep=comma,x index=0,y index=6] {csv/RBTreeCoverage.csv};
    \end{groupplot}
  \end{tikzpicture}\\
  \ref{legend}
\caption{Coverage-testing of \texttt{Data.Map.Base}, \texttt{RBTree}, and
  \texttt{XMonad.StackSet} using \toolname. Each exported function was tested
  with increasing depth limits until a single run hit a timeout ranging from one
  to thirty minutes. Lower is better for ``always-true'' and ``always-false'',
  higher is better for everything else.}\label{fig:coverage}
\end{figure}

\mypara{Results}
The results of our experiments are shown in Figure~\ref{fig:coverage}. 
Across all three libraries, \toolname achieved at least 70\% expression 
and 64\% alternative coverage at the shortest timeout of one minute per function. 
Interestingly, the coverage metrics for @RBTree@ and @Data.Map@ remain relatively constant as we increase
the timeouts, with a small jump in expression coverage between 10 and 20 minutes.
@XMonad@ on the other hand, jumps from 70\% expression and 64\% alternative
coverage with a one minute timeout, to 96\% expression and 94\% alternative
with a ten minute timeout.


There are three things to consider when examining these results. 
First is that some expressions are not evaluated due to Haskell's 
laziness (\eg the values contained in a @Map@). 
Second is that some expressions \emph{should not} be evaluated 
and some branches \emph{should not} be taken, as these only happen
when an unexpected error condition is triggered (\ie these expressions
should be dead code).
\toolname considers any inputs that trigger an uncaught exception a 
valid counterexample; the pre-conditions should rule out these inputs, 
and so we expect not to cover those expressions with \toolname.

The last remark is not intrinsically related to \toolname, 
but rather our means of collecting the coverage data. @hpc@ includes 
@otherwise@ guards in the ``always-true'' category, even though they 
cannot evaluate to anything else. 
@Data.Map@ contained 56 guards, of which 24 were marked ``always-true''. We
manually counted 21 \hbox{@otherwise@} guards, the remaining 3 ``always-true''
guards compared the size of subtrees when rebalancing to determine whether a
single or double rotation was needed; we were unable to trigger the double
rotation in these cases.
\hbox{@XMonad@} contained 9 guards, of which 4 were ``always-true''. 3 of these
were @otherwise@ guards; the remaining ``always-true'' guard dynamically checked
a function's pre-condition. If the pre-condition check had failed an error would
have been thrown by the next case, we consider it a success of \toolname that
the error branch was not triggered.

\subsection{Discussion}\label{sec:discussion}

To sum up, our experiments demonstrate that \toolname generates valid inputs:
(1) where \quickcheck fails outright, due to the low probability of
    generating random values satisfying a property;
(2) more efficiently than \lazysmallcheck, which relies on lazy
    pruning predicates; and
(3) providing high code coverage for real-world libraries with no
    hand-written test cases.


Of course our approach is not without drawbacks; we highlight five classes
of pitfalls the user may encounter.

\mypara{Laziness} in the function or in the output refinement can cause exceptions
  to go un-thrown if the output value is not fully demanded. For example,
  \toolname would decide that the result @[1, undefined]@ inhabits @[Int]@ but not
  @[Score]@, as the latter would have to evaluate @0 <= undefined < 100@. This
  limitation is not specific to our system, rather it is fundamental to any tool
  that exercises lazy programs. Furthermore, \toolname only generates
  inductively-defined values, it cannot generate infinite or cyclic structures,
  nor will the generated values ever contain $\bot$.

\mypara{Polymorphism} Like any other tool that actually runs the function under scrutiny,
  \toolname can only test monomorphic instantiations of polymorphic
  functions. For example, when testing @XMonad@ we instantiated the ``window''
  parameter to @Char@ and all other type parameters to @()@, as the properties
  we were testing only examined the window. This helped drastically reduce the
  search space, both for \toolname and \smallcheck.


\mypara{Advanced type-system features} such as GADTs and Existential types
  may prevent GHC from deriving a @Generic@ instance, which would force the
  programmer to write her own @Targetable@ instance. Though tedious, the single
  hand-written instance allows \toolname to automatically generate values
  satisfying disparate constraints, which is still an improvement over the
  generate-and-filter approach.
  
\mypara{Refinement types} are less expressive than properties written in the
  host language. If the pre-conditions are not expressible in \toolname's logic,
  the user will have to use the generate-and-filter approach, losing the benefits
  of symbolic enumeration.
  
\mypara{Input explosion} \toolname excels when the space of valid inputs is
  a sparse subset of the space of all inputs. If the input space is not
  sufficiently constrained, \toolname may spend lose its competitive advantage
  over other tools due to the overhead of using a general-purpose solver.



\section{Related Work} \label{sec:related}

\toolname is closely related to a number of lines of work on connecting
formal specifications, execution, and automated constraint-based testing. 
Next, we describe the closest lines of work on test-generation and 
situate them with respect to our approach.

\subsection{Model-based Testing}
\label{sec:model-based-testing}
Model-based testing encompasses a broad range of black-box testing tools that
facilitate generating concrete test-cases from an abstract model of the system
under test. These systems generally (though not necessarily) model the system at
a holistic level using state machines to describe the desired
behavior~\cite{DiasNeto:2007:SMT:1353673.1353681}, and may or may not provide
fully automatic test-case generation. In addition to generating test-cases, many
model-based testing tools, \eg Spec Explorer~\cite{Veanes08} will produce extra artifacts
like visualizations to help the programmer understand the model. One could view
property-based testing, including our system, as a subset of model-based testing
focusing on lower-level properties of individual functions (unit-testing),
while using the type-structure of the functions under scrutiny to provide fully
automatic generation of test-cases.

\subsection{Property-based Testing}
\label{sec:property-based-testing}
Many property-based testing tools have been developed to automatically generate
test-suites. \quickcheck~\cite{claessen_quickcheck:_2000} randomly generates
inputs based on the property under scrutiny, but requires custom generators to
consistently generate constrained inputs. \cite{Claessen14Flops} extends
\quickcheck to randomly generate constrained values from a uniform distribution.
In contrast \smallcheck~\cite{runciman_smallcheck_2008} enumerates all possible
inputs up to some depth, which allows it to check existential properties in
addition to universal properties; however, it too has difficulty generating
inputs to properties with complex pre-conditions.
\lazysmallcheck~\cite{runciman_smallcheck_2008} addresses the issue of generating
constrained inputs by taking advantage of the inherent laziness of the
property, generating \emph{partially-defined} values (\ie values
containing $\bot$) and only filling in the holes if and when they are
demanded. 
Korat~\cite{Boyapati02} instruments a programmer-supplied
@repOk@ method, which checks class invariants and method pre-conditions, to
monitor which object fields are accessed. The authors observe that unaccessed fields
cannot have had an effect on the return value of @repOk@ and are thereby able to
exclude from the search space any objects that differ only in the values of the
unaccessed fields. 
While \lazysmallcheck and Korat's reliance on functions in the 
source language for specifying properties is convenient for the 
programmer (specification and implementation in the same language), 
it makes the method less amenable to formal verification, the
properties would need to be re-specified in another language 
that is restricted enough to facilitate verification.

\subsection{Symbolic Execution and Model-checking}
\label{sec:static-analysis}
Another popular technique for automatically generating test-cases is to analyze
the source code and attempt to construct inputs that will trigger different
paths through the program. DART~\cite{DART}, CUTE~\cite{CUTE}, 
and Pex~\cite{tillmann_pexwhite_2008} all use a
combination of symbolic and dynamic execution to explore different paths through
a program. 
While executing the program they collect \emph{path predicates},
conditions that characterize a path through a program, and at the end of a run
they negate the path predicates and query a constraint solver for another
assignment of values to program variables. This enables such tools to
efficiently explore many different paths through a program, but the technique
relies on the path predicates being expressible symbolically. When the
predicates are not expressible in the logic of the constraint solver, they fall
back to the values produced by the concrete execution, at a severe loss of
precision.
Instead of trying to trigger all paths through a program, one might 
simply try to trigger erroneous behavior. 
Check 'n' Crash~\cite{jcrasher} uses the ESC/Java analyzer~\cite{ESCJava} to discover
potential bugs and constructs concrete test-cases designed to trigger 
the bugs, if they exist. Similarly, \cite{ICSE04BLAST} uses the BLAST 
model-checker to construct test-cases that bring the program to 
a state satisfying some user-provided predicate.

In contrast to these approaches, \toolname (and more generally, property-based testing) 
treats the program as a \emph{black-box} and only requires that the pre- and 
post-conditions be expressible in the solver's logic. 
Of course, by expressing specifications in the source language, 
\eg as contracts, as in PEX~\cite{tillmann_pexwhite_2008}, one can use symbolic
execution to generate tests directly from specifications.
One concrete advantage of our approach over the symbolic execution based method
of PEX is that the latter generates tests by \emph{explicitly enumerating} paths
through the contract code, which suffers from a similar combinatorial 
problem as \smallcheck and \quickcheck. In contrast, \toolname performs the 
same search \emph{symbolically} within the SMT engine, which 
performs better for larger input sizes.

\subsection{Integrating Constraint-solving and Execution}
\label{sec:constraint-solving-execution}

\toolname is one of many tools that makes specifications 
executable via constraint solving. 
An early example of this approach is 
TestEra~\cite{Marinov:2001:TNF:872023.872551} 
that uses specifications written in the Alloy 
modeling language~\cite{jackson2002alloy} to 
generate all non-isomorphic Java objects that 
satisfy method pre-conditions and class invariants. 
As the specifications are written in Alloy, one can use 
Alloy's SAT-solver based model finding to symbolically 
enumerate candidate inputs.
Check 'n' Crash uses a similar idea, and SMT 
solvers to generate inputs that satisfy a given 
JML specification~\cite{jcrasher}.
Recent systems such as SBV~\cite{sbv} and 
Kaplan~\cite{Koksal:2012:CC:2103656.2103675} 
offer a monadic API for writing SMT constraints 
within the program, and use them to synthesize 
program values at \emph{run-time}. 
SBV provides a thin DSL over the logics understood 
by SMT solvers, whereas Kaplan integrates deeply 
with Scala, allowing the use of user-defined 
recursive types and functions. 
Test generation can be viewed as a special case 
of value-synthesis, and indeed Kaplan has been 
used to generate test-suites from preconditions 
in a similar manner to \toolname.

However, in all of the above (and also symbolic execution 
based methods like PEX or JCrasher), the specifications are 
\emph{assertions} in the Floyd-Hoare sense. 
Consequently, the techniques are limited to testing 
first-order functions over monomorphic data types.
In contrast, \toolname shows how to view \emph{types} as
executable specifications, which yields several advantages.
First, we can use types to compositionally lift specifications 
about flat values (\eg @Score@) over collections (\eg @[Score]@),
without requiring special recursive predicates to describe 
such collection invariants. 
%
Second, the compositional nature of types yields a 
compositional method for generating tests, allowing 
us to use type-class machinery to generate tests for
richer structures from tests for sub-structures.
Third, (refinement) types have proven to be effective 
for \emph{verifying} correctness properties in modern
modern languages that make ubiquitous use of parametric 
polymorphism and higher order 
functions~\cite{pfenningxi98,Dunfield07,SaraswatX10,fstar,VazouRealWorld14} 
and thus, we believe \toolname's approach of making refinement types
executable is a crucial step towards %
our goal of enabling 
\emph{gradual verification} for modern languages.

\subsection*{Acknowledgements}
This work was supported by NSF grants CCF-1422471,
CNS-0964702, CNS-1223850, CCF-1218344, CCF-1018672, and
a generous gift from Microsoft Research.
We thank 
Lee Pike
and the reviewers for their 
excellent feedback on
a draft of this paper.

{
\bibliographystyle{splncs03}
\bibliography{sw}
}

\includeTechRep{
  \appendix
  \section{Instantiating \toolname Generically}\label{sec:genericapp}

\GhcGenerics defines separate types for products, data constructors, sums, and
datatypes; and uses the @TypeFamilies@ extension~\cite{Chakravarty_ATS_2005} to define an 
associated generic representation @Rep a@ for any algebraic datatype. For 
example, the standard Haskell list would be represented by the generic type
\begin{code}
  Rep [a] = C1 U1 :+: C1 (Rec0 a :*: Rec0 [a])
\end{code}
where @C1@ denotes a data constructor, @U1@ an empty product (\eg for a nullary
constructor), @:+:@ a sum, and @:*:@ a product. Additionally, @Rec0@ indicates a
reference to a user-defined type, \ie values are translated to and from the
universal representation \emph{as-needed}. We omit some of the metadata to
highlight the structural similarity between the generic representation and the
original data definition.

\ES{TODO: explain that insight of ghc-generics is that you can treat all sums
  equally and all products equally, so general approach is to define two
  type-classes: one that handles sums and another that handles products.}

\ES{TODO: explain that generic rep is tree-structured, NOT list.}


Recall that our implementation from \S~\ref{sec:list} contained three
pieces of type-specific logic, namely
(1) obtaining a list of @Ctor@s to unfold at a given depth (@ctors@),
(2) decoding a specific @Ctor@ (@decodeCtor@), and
(3) encoding a Haskell value as a logical expression (@encode@).
We now demonstrate how to generically implement these three steps for any
algebraic datatype, but first we will need two extensions to our refinement type
API, which we describe in Figure~\ref{fig:rtype-ext}.

\begin{figure}
\begin{mdframed}
\begin{CenteredBox}
\begin{code} 
ctorArity :: Ctor -> Int
mkCtor    :: Proxy (C1 c f) -> Ctor
\end{code}
\end{CenteredBox}
\end{mdframed}
\caption{Extensions to the refinement type API from Figure~\ref{fig:rtype}}\label{fig:rtype-ext}
\end{figure}

\begin{itemize}
\item{@ctorArity@} returns the number of fields that a @Ctor@ has.
\item{@mkCtor@} constructs a @Ctor@ from a \emph{proxy} for the constructor.
\end{itemize}

\subsection{Listing Constructors}\label{sec:generic-constructors}
Let us begin by writing a function @gCtors@ that will work just like @ctors@,
but for any datatype, \ie it will return a list of @Ctor@s that should be
unfolded at the given depth @d@. As is standard for \GhcGenerics we will define
a type-class for @gCtors@ and provide instances for sums, products, etc.
\GhcGenerics uses a number of different types for which we must provide class
instances, but only a few of the instances are interesting, which we show in
Figure~\ref{fig:generic-query}.
\begin{figure}[ht]
\begin{mdframed}
\begin{CenteredBox}
\begin{code}
  class GCtors f where
    gCtors :: Proxy f -> Int -> [Ctor]

  instance (GCtors f, GCtors g) => GCtors (f :+: g) where
    gCtors _ d = gCtors (Proxy :: Proxy f) d 
              ++ gCtors (Proxy :: Proxy g) d

  instance GCtors f => GCtors (C1 c f) where
    gCtors p 0
      | conArity c == 0 = [c]
      | otherwise       = []
      where
        c = mkCtor p
    gCtors p d
      = [mkCtor p]
\end{code}
\end{CenteredBox}
\end{mdframed}
\caption{Generic query generation}\label{fig:generic-query}
\end{figure}
For example, to obtain the list of @Ctor@s for a sum we simply concatenate the
lists obtained from the left- and right-hand sides of the sum.  When we reach a
specific constructor, we compare the constructor's arity with the depth; when we
reach depth 0 we only want to unfold \emph{nullary} constructors~\footnote{In
  practice one might want to do something smarter, like checking the minimum
  depth required to unfold the constructor.}.
\ES{The (C1 c f) might be confusing because the "c" was omitted in the "Rep [a]"
  example above...}
Now we can replace the call to @ctors@ in @queryList@ with
\begin{code}
  -- reproxyRep :: Proxy a -> Proxy (Rep a)
  let cs = gCtors (reproxyRep $ proxy t) d
\end{code}
\ES{i'm pretty sure (reproxyRep \$ proxy t) won't typecheck due to the existential..}
making our @query@ implementation fully datatype-generic.




  

\subsection{Decode}\label{sec:generic-decode}
Next we will tackle the process of \emph{decoding} a specific constructor from
the model. As above, we will define a type-class and show only the interesting
instances in Figure~\ref{fig:generic-decode}.
\begin{figure}[ht]
\begin{mdframed}
\begin{CenteredBox}
\begin{code}
  class GDecode f where
    gDecode :: Ctor -> [Var] -> Gen f

  instance (GDecode f, GDecode g) => GDecode (f :+: g) where
    gDecode c vs =  L1 <$> gDecode c vs
                <|> R1 <$> gDecode c vs

  instance GDecodeFields f => GDecode (C1 c f) where
    gDecode c vs 
      | c == mkCtor (Proxy :: Proxy (C1 c f))
      = C1 . snd <$> gDecodeFields vs
      | otherwise
      = empty

  class GDecodeFields f where
    gDecodeFields :: [Var] -> Gen ([Var], f)

  instance Targetable a => GDecodeFields (Rec0 a) where
    gDecodeFields (v:vs) = do
      x <- decode v
      return (vs, Rec0 x)
\end{code}
\end{CenteredBox}
\end{mdframed}
\caption{Generic decoding of Haskell values}\label{fig:generic-decode}
\end{figure}

Given an arbitrary sum, we do not know whether the constructor we are looking
for is in the left or right sub-sum, so we must try both.
Once we reach an individual constructor, we can check whether it is the correct
constructor using the forementioned @mkCtor@ function. If the check is
successful, we can go ahead and decode the constructor's fields using
@gDecodeFields@ and wrap them up, otherwise we signal that the next element of
the sum should be tried.

@gDecodeFields@ comes from an auxiliary type-class that we use to decode the
fields of a product. As \GhcGenerics represents sums and products as
\emph{trees} instead of lists, we have @gDecodeFields@ return the list of
@Var@s that still need to be decoded in addition to the decoded value.
Again, most of the instances are uninteresting and simply involve traversing the
product while decoding each field. The interesting instance arises when we want
to encode an individual field. Recall that products are represented with
recursive references to the user-defined type, \eg @Rec0 [a]@. So when we reach
an individual field, we will have to decode a value of the \emph{user-defined}
type.

We can now replace the @decodeCtor c fs'@ in our original implementation of @decode@
with
\begin{code}
  -- to :: Generic a => Rep a -> a
  to <$> gDecode c xs
\end{code}
%
%
%
%
\subsection{Check}\label{sec:generic-check}
Finally, let us examine how to generically \emph{check} that Haskell value 
inhabits a refinement type with the type-class in Figure~\ref{fig:generic-check}.
\begin{figure}[ht]
\begin{mdframed}
\begin{CenteredBox}
\begin{code}
  class GCheck f where
    gCheck :: f -> RefType -> Gen (Bool,Var)
    
  class GCheckFields f where
    gCheckFields :: f -> [(Var, RefType)]
                 -> SMT (Bool, [Var], [(Var, RefType)])
                 
  instance GCheckFields f => GCheck (C1 c f) where
    gCheck (C1 f) t = do
      let c       = mkCtor (Proxy :: Proxy (C1 c f))
      let fts     = unfold c t
      (b, vs, _) <- gCheckFields f fts
      v          <- apply c vs
      let t'      = t `subst` [(binder t, v)]
      b'         <- eval t'
      return (b', v)
      
  instance Targetable a => GCheckFields (Rec0 a) where
    gCheckFields (Rec0 a) ((f, t) : fts) = do
      (b, v)  <- check a t
      let fts' = fts `subst` [(f, v)]
      return (b, [v], fts')
\end{code}
\end{CenteredBox}
\end{mdframed}
\caption{Generic checking of Haskell values against refinement types.}\label{fig:generic-check}
\end{figure}

Checking a sum just involves stripping away levels of indirection until we reach
the actual constructor, at which point we need to unfold the constructor and
check its fields. We then @apply@ the constructor to the resulting symbolic
values, @subst@itute the resulting @Refinement@ for @t@'s @binder@ and
@eval@uate the result.

@gCheckFields@ checks the fields of a product, and is itself a type-class method.
As with @gDecodeFields@ the only interesting instance of @GCheckFields@ deals
with checking an individual field, where we have a value of the user-defined
type and must use the original @check@ method.

Now we can provide a default implementation for @check@
\begin{code}
  -- from :: Generic a => a -> Rep a
  check v t = gCheck (from v) t
\end{code}
thus replacing the last bit of type-specific logic in our @Targetable@
implementation.






}

\end{document}